# Charge transfer gap for *T'*-*RE*$_2$CuO$_4$ and *T*-La$_2$CuO$_4$ as estimated from Madelung potential calculations

A. Tsukada [a, b, *], H. Shibata [b], M. Noda [b, c], H. Yamamoto [b], M. Naito [a]

[a] Department of Applied Physics, Tokyo University of Agriculture and Technology, 2-24-16 Naka-cho, Koganei, Tokyo 184-8588, Japan

[b] NTT Basic Research Laboratories, NTT Corporation, 3-1 Morinosato-Wakamiya, Atsugi, Kanagawa 243-0198, Japan

[c] Tokyo University of Science, 2641 Yamazaki, Noda, Chiba 278-8510, Japan

## Abstract

*T'*-*RE*$_2$CuO$_4$ (*RE*: rare-earth element), after appropriate "reduction", has fairly high conductivity and also exhibits clear Fermi edge in photoemission spectroscopy. To clarify the origin of conductivity in the *T'* mother compounds, we evaluated the unscreened charge-transfer gap ($\Delta_0$) for *T'*-*RE*$_2$CuO$_4$ and *T*-La$_2$CuO$_4$. The $\Delta_0$ value for *T'*-compounds almost linearly decreases with increasing the ionic radius of *RE* from 12.24 eV for *T'*-Tm$_2$CuO$_4$ to 9.90 eV for *T'*-La$_2$CuO$_4$. The results qualitatively explain metallic conductivity in *T'*-*RE*$_2$CuO$_4$ for large *RE*.

*Corresponding author.

Dr. Akio Tsukada

Postal address: Naito Laboratory, Department of Applied Physics, Tokyo University of Agriculture and Technology, 2-24-16 Naka-cho, Koganei, Tokyo 184-8588, Japan

Phone: +81-42-388-7229

Fax: +81-42-385-6255

E-mail address: atsukada@cc.tuat.ac.jp

## 1. Introduction

It has been believed for a long time that high-$T_c$ superconductivity (HTSC) develops in a Mott-insulating mother compound by doping either holes or electrons. This picture is called as the “doped Mott-insulator scenario” for HTSC. Recently, however, we discovered superconductivity with $T_c$ > 20 K in *nominally* “undoped” $La^{3+}_{2-x}RE^{3+}_{x}CuO_4$ (*RE* = rare-earth elements) with the $Nd_2CuO_4$ (*T’*) structure [1-3], which has raised strong skepticism to the doped Mott-insulator scenario for HTSC. Before this discovery, we had already noticed that $T'$-$RE_2CuO_4$, after appropriate “reduction”, has fairly high conductivity [4, 5] and also exhibits clear Fermi edge in our *in-situ* photoemission spectroscopy with thin-film specimens prepared by MBE. The conductivity of $T'$-$RE_2CuO_4$ monotonically increases with increasing the ionic radius ($r_i$) of *RE* or equivalently increasing the in-plane lattice constant ($a_0$) of $T'$-$RE_2CuO_4$: resistivity at room temperature ($\rho_{300\,K}$) varies from ~ 1 Ωcm for $T'$-$Tb_2CuO_4$ [$r_i(Tb^{3+})$ = 1.040 Å, $a_0$ = 3.892 Å] to ~ 0.002 Ωcm for $T'$-$La_2CuO_4$ [$r_i(La^{3+})$ = 1.160 Å, $a_0$ = 4.026 Å]. Furthermore, metallic behavior is observed down to 150 K for large *RE* ions such as La, Pr, and Nd [1]. The behavior is in a very sharp contrast to that of $La_2CuO_4$ with the $K_2NiF_4$ (*T*) structure, which is highly insulating ($\rho_{300\,K}$ ~ 100 Ωcm) [4]. Between *T*- and $T'$-$La_2CuO_4$, $\rho_{300\,K}$ differs by five orders of magnitude, and the difference becomes much larger at lower temperature.

With regard to conductivity in oxides, among hundreds of oxides, including high-$T_c$ cuprates, the majority of undoped oxides are insulating, but some exhibit metallic conductivity. Torrance *et al.* [6] gave a successful explanation for the large difference in conductivity among oxides based on the ionic model developed by Zaanen, Sawatzky, and Allen. In this Zaanen-Sawatzky-Allen framework, each oxide can be characterized by three parameters: Coulomb correlation energy ($U$), the charge transfer energy ($\Delta$), and the bandwidth ($W$). Insulating oxides are classified in two types: Mott-Hubbard insulator ($\Delta > U$) and charge-transfer (CT) insulator ($U > \Delta$). Torrance *et al.* evaluated "unscreened" Coulomb correlation energy ($U_0$) and charge-transfer energy ($\Delta_0$) for many oxides using electrostatic Madelung potentials, empirical ionization energies of cations, and electron affinity of $O^-$. Then they proposed a semi-quantitative criterion for $U_0$ and $\Delta_0$ to distinguish metallic and insulating oxides. Namely, oxides with $\Delta_0 < 10$ eV or $U_0 < 11$ eV are metallic, while oxides with $\Delta_0 > 10$ eV and $U_0 > 11$ eV are insulating, and oxides showing metal-to-insulator transition are located on the boundary. In this work, we evaluated the $\Delta_0$ for $T'$-$RE_2CuO_4$ and $T$-$La_2CuO_4$, based on the formulation proposed by Torrance *et al.* [6] The results qualitatively explain metallic conductivity in $T'$-$RE_2CuO_4$ for large $RE$ and also a large difference in conductivity between $T'$-$La_2CuO_4$ and $T$-$La_2CuO_4$.

## 2. Formulation

The mother compounds of high-$T_c$ cuprates can be regarded as CT insulators, whose energy gap $\Delta$ may be given by the energy-level difference between the occupied O2$p$ orbital and the unoccupied Cu3$d$ orbital. In the framework of the ionic model, the unscreened $\Delta_0$ is given by

$$\Delta_0 = e\Delta V_M + A(O^-) - I_2(Cu^+) - e^2/d_{Cu\text{-}O}, \qquad (1)$$

where $\Delta V_M$ is the difference in electrostatic Madelung site potentials between in-plane oxygen and copper ($\Delta V_M \equiv V_M(O_{pl}) - V_M(Cu)$), $A(O^-)$ the electron affinity of $O^-$, $I_2(Cu^+)$ the ionization potential of $Cu^+$, and the term $e^2/d_{Cu\text{-}O}$ the Coulomb attraction between the excited electron and the hole. The values of $A(O^-)$ and $I_2(Cu^+)$ are structure independent, and -7.70 and 20.29 eV were used for $A(O^-)$ and $I_2(Cu^+)$, respectively. The values of $e\Delta V_M$ and $e^2/d_{Cu\text{-}O}$ are calculated with parameters ($a_0$, $c_0$, and $d_{Cu\text{-}O} = a_0/2$) listed in Table 1. The positions of each atom in the structure were taken from ref. 7 for *T'* compounds and ref. 8 for *T*. In the calculation of *T'* compounds, the atomic positions are available only for $Nd_2CuO_4$, so, for all other $RE_2CuO_4$, the same relative atomic positions as $Nd_2CuO_4$ were used and only the lattice parameters were varied [9]. The Madelung potential calculations were performed by the program developed by K. Kato and F. Izumi [10].

## 3. Results and discussion

The calculated results are summarized in Table 1. First, we compare *T'*-$La_2CuO_4$ and *T*-$La_2CuO_4$. The $\Delta_0$ values are significantly different: 9.90 eV for *T'*-$La_2CuO_4$ and 13.69 eV for *T*-$La_2CuO_4$. The difference mostly comes from the difference in $eV_M$(Cu) since there is only a small difference (0.19 eV) in $eV_M(O_{pl})$ between the two compounds. The significantly smaller absolute value $|eV_M(Cu)|$ in *T'*-$La_2CuO_4$ than in *T*-$La_2CuO_4$ is primarily due to the absence of apical oxygen and secondly due to the substantially longer Cu-$O_{pl}$ distance, both of which reduce the electrostatic potential at the copper site. As a result, the Cu3*d* energy level ($3d^9$) in *T'*-$La_2CuO_4$ is lower than in *T*-$La_2CuO_4$ whereas the O2*p* level is almost unchanged in the two compounds.

Next, we compare the $\Delta_0$ values of *T'*-$RE_2CuO_4$ with different *RE*. Figure 1 plots $\Delta_0$ as a function of the ionic radius ($r_i$) of *RE* in *T'* compounds. The $\Delta_0$ value almost linearly decreases with increasing $r_i$ or equivalently the lattice constants ($a_0$ and $c_0$) from 12.24 eV for *T'*-$Tm_2CuO_4$ to 9.90 eV for *T'*-$La_2CuO_4$. In this case, both the absolute values, $|eV_M(Cu)|$ and $|eV_M(O_{pl})|$, decreases linearly with lattice expansion.

One must remember that the above $\Delta_0$ values are obtained by neglecting any of the effects of the overlap between ions (covalency, hybridization, screening, electronic polarizability, band width, etc.). As stressed by Torrance *et al.*, the basic assumption is not that these effects are small, rather that they are similar among a certain class of oxides: namely, the *difference* in actual $\Delta$ is dominated by the *difference* in unscreened $\Delta_0$. Next we compare the calculated $\Delta_0$ with the value ($\Delta_{CT}^{exp}$) experimentally obtained. The observed $\Delta_{CT}^{exp}$ is ~ 2.0 eV for *T*-$La_2CuO_4$ [11], and $\Delta_{CT}^{exp}$ of *T'*-$RE_2CuO_4$ ranges from 1.5 eV (*RE* = Pr) to 1.7 eV (*RE* = Gd) [12]. The $\Delta_{CT}^{exp}$ values among all the *T* and *T'* cuprates are within a narrow range of 1.5 eV to 2.0 eV, depending only very weakly on the Cu-$O_{pl}$ bond length. This is in apparent contradiction to the above calculations. We speculate the reason for this contradiction as follows. In general, *T'* cuprates lose conductivity significantly if they are oxidized. For example, ozone oxidation increases $\rho_{300\ K}$ from ~ 0.002 Ωcm for *T'*-$La_2CuO_4$ to ~ 1 Ωcm for *T'*-$La_2CuO_{4+\delta}$. This is because impurity oxygen atoms at the apical site act as strong scattering centers (as well as pair breakers) [13]. Hence it is very important to clean up impurity oxygen atoms as much as possible in order to see the generic properties of *T'* cuprates. Arima *et al.* carried out careful examinations of the *reduction* effects on optical spectra of *T'*-$(Pr,Ce)_2CuO_4$ [14]. They reported that reduction enhances the infrared reflectivity as well as diminishes the peak intensity of the optical conductivity that was assigned as due to the CT gap transition. They attributed these changes to net electron doping due to oxygen deficiency. However, as noted by themselves, *reduction* causes essentially no observable change in the lattice constants. If electron doping *were* actually achieved, it *would* result in Cu-O bond stretching. Instead, as claimed by Moran *et al.* [15] and Tarascon *et al.* [16] based on chemical analysis, as-prepared *T'* cuprates contain a fair amount of impurity oxygen atoms at the apical site, which is also supported by neutron diffraction studies by Radaelli *et al* [7]. Hence, we believe that the $\Delta_{CT}^{exp}$ reported thus far for *T'* cuprates is not an intrinsic value, but a value altered by the impurity-oxygen effect. We estimate the shift in $\Delta_0$ due to adding an $O^-$ ion at the apical site and reducing the charge at the neighboring four oxygen atoms (O(2)) in the $RE_2O_2$ plane from $O^{-2}$ to $O^{-1.75}$ (assumption of peroxide formation). The increase in $\Delta_0$ amounts to more than 1 eV per one apical oxygen. Energy shift of such order may account for the discrepancy between our calculations and the past experiments on *T'* cuprates.

Finally we mention the indications of our Madelung potential calculations. The unscreened $\Delta_0$ of *T'*-$La_2CuO_4$ is 9.9 eV, which is the smallest in high-$T_c$ cuprates. According to Torrance's criterion, this value is located on the boundary ($\Delta_0^{cr}$ = 10 eV) that separates metallic and insulating oxides. The $\Delta_0$ values for *T'*-$Pr_2CuO_4$ and *T'*-$Nd_2CuO_4$ are 10.88 eV and 11.04 eV, which are just above $\Delta_0^{cr}$. This indicates that the charge transfer gap of *T'*-$RE_2CuO_4$ should be very small or may even close for large *RE*, especially La, which is actually consistent with the metallic resistivity observed for these compounds. The most crucial test to confirm this statement is optics measurements on "*well-reduced*" films, and such studies are under way.

## 4. Summary

We have calculated the unscreened charge-transfer gap $\Delta_0$ of *T'*-*RE*$_2$CuO$_4$ and *T*-La$_2$CuO$_4$ based on the ionic model. The results indicates that $\Delta_0$ of *T'*-*RE*$_2$CuO$_4$ is, in general, smaller than $\Delta_0$ of *T*-La$_2$CuO$_4$, which is due to the absence of apical oxygen and the longer Cu-O bond length in *T'* cuprates. Especially $\Delta_0$ of *T'*-La$_2$CuO$_4$ is as small as 9.9 eV, and is located on the metal-insulator boundary of oxides. This indicates that the charge transfer gap of *T'*-La$_2$CuO$_4$ should be very small or even may close. Finally, for such low-$\Delta_0$ metal, the ionic picture loses its meaning, and one has to resort the band picture.

**Acknowledgements**

The authors thank Dr. T. Yamada, Dr. H. Sato, Dr. S. Karimoto, Dr. K. Ueda, and Dr. A. Matsuda for helpful discussions, and Dr. T. Makimoto for his support and encouragement.

Table 1 $a_0$, $c_0$, calculated $eV_M$(O), $eV_M$(Cu), $e\Delta V_M$, and $\Delta_0$ of $T'$-$RE_2CuO_4$ and $T$-$La_2CuO_4$.

| Material | $a_0$ (Å) | $c_0$ (Å) | $eV_M(O_{pl})$ | $eV_M$(Cu) | $e\Delta V_M$ (eV) | $\Delta_0$ (eV) |
|---|---|---|---|---|---|---|
| $T'$-$La_2CuO_4$ | 4.026 | 12.550 | 21.82 | -22.65 | 44.47 | 9.90 |
| $T'$-$Pr_2CuO_4$ | 3.958 | 12.288 | 22.19 | -23.05 | 45.24 | 10.88 |
| $T'$-$Nd_2CuO_4$ | 3 .943 | 12.177 | 22.30 | -23.14 | 45.44 | 11.04 |
| $T'$-$Sm_2CuO_4$ | 3.905 | 11.929 | 22.54 | -23.38 | 45.92 | 11.46 |
| $T'$-$Eu_2CuO_4$ | 3.894 | 11.882 | 22.61 | -23.45 | 46.06 | 11.57 |
| $T'$-$Gd_2CuO_4$ | 3.888 | 11.859 | 22.65 | -23.48 | 46.13 | 11.63 |
| $T'$-$Tb_2CuO_4$ | 3.880 | 11.815 | 22.70 | -23.53 | 46.23 | 11.71 |
| $T'$-$Dy_2CuO_4$ | 3.869 | 11.771 | 22.76 | -23.60 | 46.36 | 11.83 |
| $T'$-$Ho_2CuO_4$ | 3.861 | 11.721 | 22.81 | -23.65 | 46.46 | 11.91 |
| $T'$-$Er_2CuO_4$ | 3.840 | 11.637 | 22.94 | -23.78 | 46.72 | 12.13 |
| $T'$-$Tm_2CuO_4$ | 3.830 | 11.578 | 23.01 | -23.84 | 46.85 | 12.24 |
| $T'$-$Y_2CuO_4$ | 3.861 | 11.721 | 22.81 | -23.65 | 46.46 | 11.91 |
| $T$-$La_2CuO_4$ | 3.803 | 13.150 | 21.63 | -26.72 | 48.35 | 13.69 |

**Figure caption**

Fig. 1 Variation of calculated charge-transfer gap ($\Delta_{CT}^{cal}$) for *T'*-$RE_2CuO_4$ as a function of ionic radius ($r_i$) of *RE*.

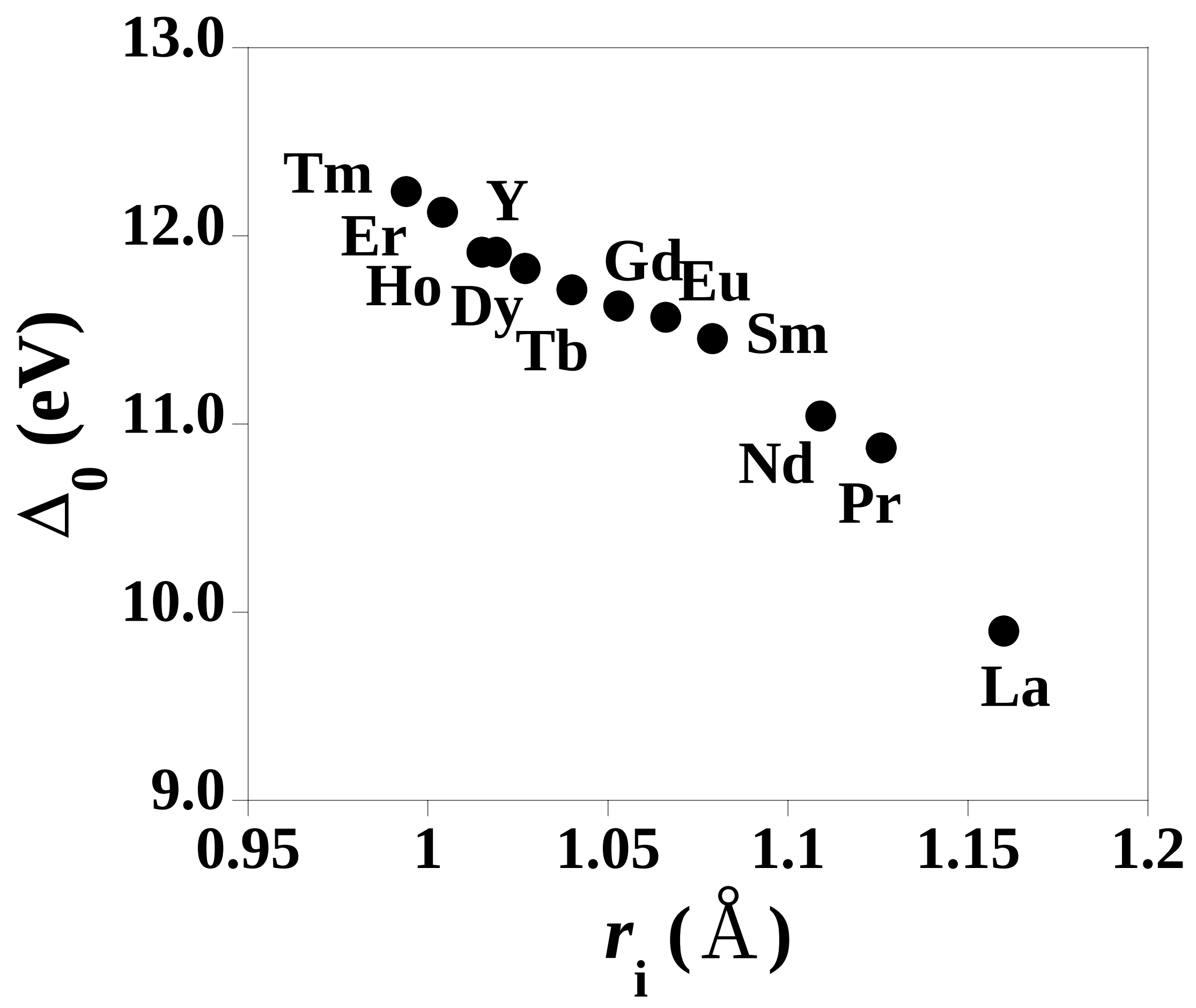

**Fig. 1, PCP-9**